# Ambipolar Ion Pumping with Ratchet-Driven Active Membranes

Alon Herman[1] and Gideon Segev[1*]

[1]School of Electrical Engineering, Tel Aviv University, Tel Aviv 6997801, Israel

[*]email: gideons1@tauex.tau.ac.il

## ABSTRACT

In recent years there has been significant progress in the development of artificial ion pumping membranes. Ion pumps based on asymmetric nano-pores have been shown to operate as ionic current rectifiers, thus pumping a net ion flux against a concentration gradient even when driven with unbiased ac signals. However, since ion transport relies on charged nano-pores, it is selective to either cations or anions, and thus cannot pump both cations and anions simultaneously. In this paper, we present a model for an electronically active membrane which is based on a *flashing* ratchet mechanism. The model includes adjacent electrolyte reservoirs and ion-ion interactions, which were not accounted for in prior similar models, and thus provides a better understanding of the driving mechanism and potential capabilities and limitations. It is shown that unlike most other proposed ion pumps, the ratchet-based ion pump (RBIP) drives both cations and anions in the same direction and up a concentration gradient. This process, referred to as *ambipolar* ion pumping, is shown to be highly robust for many electrolyte compositions and input signals. The membrane is composed of alternating conductive thin layers (electrodes), separated by insulating layers in an asymmetric design. With insulating layers thickness of 70 and 30 nm and an input signal amplitude of 0.5 V, the device drives a salt flux of 0.03 mol/m²s in a mildly saline solution (10 mM). Thus, RBIPs may pave the way for many exciting future applications involving ambipolar ion pumping, most notably for desalination.

# I. INTRODUCTION

Ion pumps can be found in any living cell membrane and are essential to many biologic processes [1,2]. Because of their wide applicability, various approaches were taken towards the realization of artificial ion pumps [3]. For example, devices that pump ions with loading-unloading cycles using chemically gated membranes [4,5], and light driven proton pumping with photoacid-dye- sensitized membranes [6,7], or membranes doped with spiropyrans [8]. Ion pumps based on membranes with spatially asymmetric nano-pores (such as conical nano-pores) have gained significant interest in recent years [9–12]. In these membranes, surface-charged nano-pores function as ionic current rectifiers, and the transmembrane potential is controlled by external electrodes. When the transmembrane potential is alternated, the rectified ion transport results in ion pumping even if the input signal is unbiased. Since the nano-pores walls are charged, there is an excess of counter-ions in the channel, making the membrane selective for either cation or anion transport. Gated nano-porous membranes can potentially be used for *ambipolar* ion pumping (i.e., driving both cations and anions in the same direction) or water pumping by alternating their surface charge and therefore their ion selectivity [13–16].

An electric ratchet is a device that utilizes a temporal modulation of a spatially asymmetric electric potential to drive a steady state flux of charged particles [17,18]. Ion pumps based on asymmetric nano-pores could be viewed as *tilting* (or *rocking)* ratchets, where the asymmetric potential profile within the nano-pores is 'tilted' by an externally applied electric field [10,19]. As such, they are driven with external electrodes and the membrane itself does not provide a driving force for charge transport. In *flashing* ratchets, the potential profile within the nano-pores is directly controlled and does not require external driving electrodes. Moreover, since potential gradients are applied to electrodes embedded within the membrane at short distances from each other (compared to the Debye length), it is possible to maintain high electric fields along the nano-channels with minimal ohmic losses and without energetically expensive electrochemical reactions.

The use of flashing ratchets for ion transport was only recently suggested theoretically [20–22] and demonstrated experimentally [23]. Theoretical studies have shown that flashing ratchets have unique capabilities for tunable ion-ion selectivity [21,22], and potentially a higher energetic efficiency than reverse osmosis for nanofiltration [20]. Prior modeling of ion pumping flashing ratchet systems assumed an ideal system in the sense that the potential distribution was predetermined, and ion-ion interactions were not taken into consideration. Furthermore, fluctuations in the reservoirs of the electric potential and ion concentration were not accounted for [20,24]. Ratschow et al. simulated a gated conical nanopore with alternating bias [16]. Under some conditions this device can be described as a flashing ratchet, however, the authors focused on water flow through the device and many questions related to its ion pumping performance remained unanswered.

Here we present a computational study on the performance of an electrically active membrane, driven with a flashing ratchet mechanism in an electrolytic environment. The simulation accounts for the essential transport mechanisms, the coulombic interaction between ions and all reservoir effects. The suggested membrane structure consists of alternating electrically conductive and insulating thin layers. The possibility of fabricating such multi-layered structures using conventional thin film deposition techniques provides great promise for achieving different electric potential landscapes within a solution. The specific design presented here, which creates a sawtooth-shaped potential, is the simplest ratchet profile that could be realized with the layered design. However, this structure could plausibly be expanded with additional layers to create a more complex ratchet class, called *reversible* ratchet, that could potentially transport ions at a very high efficiency [25].

## II. SIMULATION DETAILS

FIG. 1(a) shows a schematic illustration of an RBIP membrane placed between two electrolyte reservoirs. The RBIP is constructed from three thin electrically conductive layers through which a time-dependent voltage signal, $V_{in}(t)$, is applied, according to the scheme. Separating the conducting layers, are two insulating layers with different thicknesses, thus when a voltage is applied, an asymmetric potential distribution within the nano-channel is obtained. To reduce the computational burden, the RBIP performance is simulated using a one-dimensional model, as shown in FIG. 1(b). This simplified model can describe the device performance adequately under the following assumptions: (i) the diameter of the nano-channels is similar to the Debye length in the electrolyte (i.e., minimal potential decay towards the center of the nano-channel), (ii) the density of the nano-channels in the membrane is sufficiently large, so that the effective reservoirs diameter is close to the nano-channel diameter. The simplified model also serves as a first step that provides basic mechanistic information prior to examining a more complex three-dimensional model. It is also assumed that that the nano-channels are uncharged, and the conducting layers (electrodes) are ideally polarizable (no Faradaic reactions). Lastly, solvent convection is not considered. Although electroosmotic (solvent) flux is largely constrained for small nano-channels (assuming no slip at the nano-channels walls) [26], it is difficult to predict its effects on the device's performance, especially at high voltages and small inter-electrode distance. Therefore, a more detailed account of convection effects (by coupling the Navier-Stokes equations) is left for future work.

The RBIP device has a length $L$, with a symmetry factor $x_c$, and the width of the electrolyte reservoirs is $W$ (assumed much larger than the Debye length). The electrodes are modeled as points (0D). The distribution of the electric potential, and of the cations and anions (and their ionic current), in response to an input signal are found by solving the one-dimensional Nernst – Planck – Poisson equations using COMSOL Multiphysics® v6.1. Assuming a symmetric binary $z:z$ electrolyte, no solvent convection, and no bulk reactions, the governing equations are:

$$-\varepsilon \frac{\partial^2 \phi}{\partial x^2} = Fz(C_+ - C_-), \tag{1}$$

$$\frac{\partial C_\pm}{\partial t} = -\frac{\partial J_\pm}{\partial x} = -\frac{\partial}{\partial x}\left(-D_\pm \frac{\partial C_\pm}{\partial x} \mp \frac{zD_\pm}{V_{th}} C_\pm \frac{\partial \phi}{\partial x}\right), \tag{2}$$

Here, $\phi$ is the electrical potential in the solution, $z$ is the ion valence, $V_{th} = RT_r/F$ is the thermal voltage, and $C_+$ and $C_-$ are the cations and anions concentrations respectively. $J_\pm$, and $D_\pm$ are the cations and anions flux and diffusion coefficient, respectively. $\varepsilon$, $F$, $R$, and $T_r$ are the dielectric permittivity of the solvent, Faraday's constant, universal gas constant, and the temperature, respectively.

The boundary conditions of the potential determine how the ratchet is electrically driven. The input signal is applied to the middle electrode, relative to the left and right electrodes. In addition, A reference point for the potential is defined at the left edge of the system ($x = 0$), and the right edge of the system is defined as electrically insulating. Since the signal is electrically floating (with respect to the bulk solution in both reservoirs), there is no EDL formed at the reference point. Hence, the boundary conditions for the potential can be formulated as:

$$\phi(x = W + x_c L, t) - \phi(x = W, t) = V_{in}(t), \tag{3a}$$

$$\phi(x = W, t) = \phi(x = W + L, t), \tag{3b}$$

$$\phi(x = 0, t) = 0, \tag{3c}$$

$$\partial\phi/\partial x\, (x = 2W + L, t) = 0. \tag{3d}$$

More information on the electric potential boundary conditions is given in the Supplemental Material [27]. The input signal $V_{in}(t)$ is a periodic square wave signal that is defined for some integer time-period, $n$, as follows:

$$V_{in}(t) = \begin{cases} V_{Max}, & nT < t < (n+\delta)T \\ \alpha V_{Max}, & (n+\delta)T < t < (n+1)T \end{cases}. \tag{4}$$

Each time-period is described by a duty-cycle $\delta = \tau_1/T$, which is the ratio between the time-duration of the first step, $\tau_1$, where the potential is at its maximum value $V_{Max}$, to the total period duration $T (= 1/f)$. The time-period is completed with a second step, in which the potential is multiplied by a potential symmetry factor in the range $-1 \leq \alpha \leq 0$. The variable $\tau = t - nT$ is defined within each time-period and is set to zero at the beginning of each period.

Electroneutrality is maintained at the outer edges of the reservoirs since $W$ is much larger than the Debye length. Thus, we can define the ion concentration at the outer edge of the left reservoir as $C_L(t) = C_\pm(x = 0, t)$, and at the outer edge of the right reservoir as $C_R(t) = C_\pm(x = 2W + L, t)$. The net ion flux at some time-period $n$, is defined as:

$$J_{\pm,net}(x, nT) = (1/T) \int_0^T J_\pm(x, nT + \tau)\, d\tau. \tag{5}$$

For convenience some of the variables are normalized, such that: $\tilde{t} = t/T$, $\tilde{\tau} = \tau/T$, $\tilde{x} = x/L$, $\tilde{C} = C/C_0$, and $\tilde{J}_\pm = J_\pm/(D_\pm C_0/W)$, where $C_0$ is the bulk electrolyte concentration.

Two types of systems are simulated. The first is a *closed system* where the ionic fluxes are set to zero at the reservoir outer edges, i.e., $J_\pm(x = 0, t) = J_\pm(x = 2W + L, t) = 0$. The initial conditions are: $\phi(x, 0) = 0$, and $C_\pm(x, t = 0) = C_0$. Since the flux is zero at the outer edges of the domain, the net ion flux between the reservoirs was calculated according to the change in concentration at the outer edges. More details on the net flux calculation in the closed system can be found in the Supplemental Material [27]. Each simulation is run until the per-period average ion concentration at the reservoir outer edges reaches a steady state. At this point the average ion flux (over a time-period) is zero. Since the total amount of ions in the system is constant, this type of system allows us to follow changes in ion concentrations between the reservoirs over time. In the second type of system, termed *open system*, a constant bulk concentration boundary condition is set at the reservoir edges, such that $C_L = C_0$, and $C_R \geq C_L$. Each simulation is run until the per-period average ion flux at the reservoirs outer edges reaches a steady state. Due to the constant concentrations at the outer edges, a non-zero steady state flux can be achieved, and the pumping performance of the device can be analyzed as a function of $C_R/C_L$. The initial conditions for the open system and additional simulation definitions are given in the Supplemental Material [27].

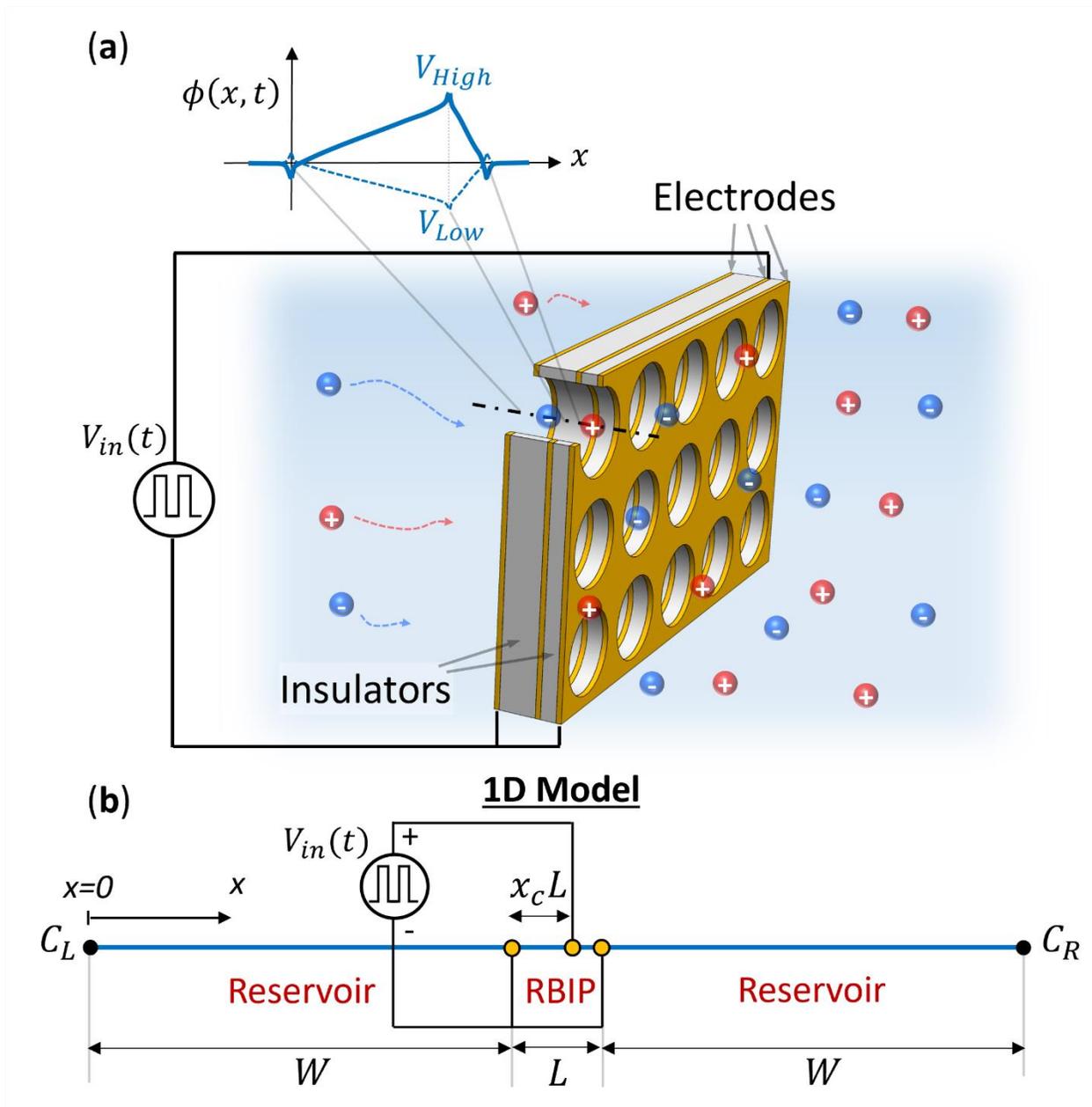

FIG. 1. (a) An illustration of the RBIP membrane, acting between two electrolyte reservoirs. The device is constructed from three thin electrically conducting layers (electrodes), through which a time-dependent voltage signal $V_{in}(t)$ is applied, and two insulating layers between them with different thicknesses. Thus, an asymmetric potential distribution is induced within the nano-channel. (b) A one-dimensional model of the system.

## III. RESULTS AND DISCUSSION

### A. RBIP operation in a *closed system*

FIG. 2 shows the RBIP operation in a closed system. The ratchet has a spatial period $L = 100$ nm, symmetry factor $x_C = 0.7$, and the width of the reservoirs is $W = 5L$. The initial concentration is $C_0 = 1$ mM (which corresponds to Debye length $\lambda_D \cong 10$ nm, at room temperature). The input $V_{in}(t)$ is a symmetric square wave signal, with: $V_{Max} = 0.5$ V, $\alpha = -1$, $\delta = 0.5$, and $f = 1$ MHz. FIG. 2(a) shows the net salt flux ($J_+ = J_-$) and the changes over time of the ion concentration at the right and left edges of the reservoirs. The electrolyte is symmetric with diffusion coefficients of $D_\pm = 2 \cdot 10^{-9}$ m²/s, and ion valence $z = 1$. FIG. 2(b) shows the same for NaCl solution, which corresponds to $D_{Na^+} = 1.33 \cdot 10^{-9}$ m²/s and $D_{Cl^-} = 2.03 \cdot 10^{-9}$ m²/s. In both cases the net transport is ambipolar, i.e., the ratchet is driving cations and anions in the same direction. Since the flux is positive, the ion concentration in the left reservoir decreases, while the concentration in the right reservoir increases. The concentration gradient that is built between the reservoirs opposes the ratchet operation, until finally at $\tilde{t} \sim 300$ the concentration at both edges reaches a steady state. The net flux reaches a maximum at $\tilde{t} = 24$ for the symmetric electrolyte and $\tilde{t} = 29$ for NaCl (FIG. 2(a) and FIG. 2(b) respectively), and then gradually declines until it reaches zero at $\tilde{t} \sim 300$ when the ratchet drive is balanced by the concentration gradient between the reservoirs. The overall behavior and steady state concentration ratio $C_R/C_L$ of the symmetric electrolyte and NaCl is very similar. However, in NaCl, sodium has a lower diffusion coefficient than chloride, and as a result, Coulombic interaction will impede the chloride transport. This leads to a lower salt flux and slightly longer time to reach steady state (FIG. 2(b)). Due to a net accumulation of ions in the ratchet area, the total number of ions that are added to the right reservoir is slightly smaller than the number of ions that are extracted from the left reservoir (FIG. 2(a)-(b)). This effect becomes negligible for larger reservoirs volumes. FIG. 2(c) shows the spatial concentration distribution of the cations (dot-dash red line) and anions (solid blue line), for $D_\pm = 2 \cdot 10^{-9}$ m²/s (same conditions as FIG. 2(a)). The region inside the membrane ($5 \cdot 10^{-7} < x < 6 \cdot 10^{-7}$ m) is whited-out for visual clarity. The distributions are shown right before the switch to a positive potential ($\tilde{t} = 0^-$), for different time-periods along the process. At the initial state ($\tilde{t} = 0$), the concentration of the ions is uniform everywhere in the domain, thus $C_\pm = C_0$. At $\tilde{t} = 24$, where the flux is maximal, the ratchet extracts ions from the left reservoir and transports them to the right reservoir. Eventually the system reaches a steady concentration difference between the reservoirs, as shown for $\tilde{t} = 500$. At this point the net flux is zero since the ratchet operation is completely negated by the concentration difference. Although not trivial, net ion transport is obtained here with an unbiased potential input (i.e., $\alpha = -1, \delta = 0.5$), which is in-line with previous findings of Brownian pumps [28].

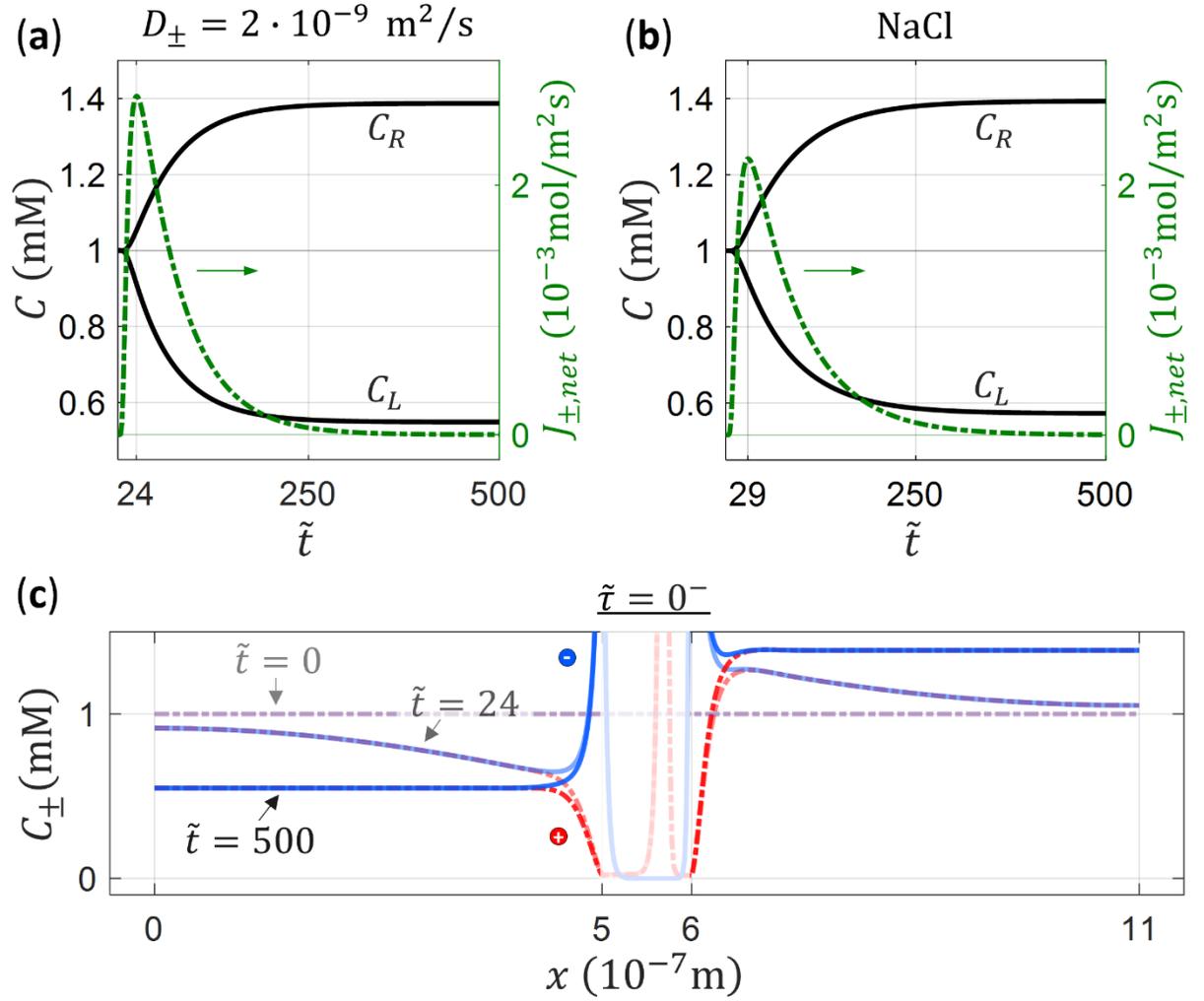

FIG. 2. RBIP operation in a *closed system*. Ion concentration over time at the right and left edges of the reservoirs (filled black lines, left y-axis), and the net salt flux (dot-dash green line, right y-axis), for a binary electrolyte ($z = 1$) with (a) symmetric diffusion coefficients, $D_\pm = 2 \cdot 10^{-9}$ m²/s, and (b) NaCl ($D_{Na^+} = 1.33 \cdot 10^{-9}$ m²/s and $D_{Cl^-} = 2.03 \cdot 10^{-9}$ m²/s). (c) Spatial concentration distribution of the cations (dot-dash red line) and anions (solid blue line) for $D_\pm = 2 \cdot 10^{-9}$ m²/s. The region inside the membrane ($5 \cdot 10^{-7} < x < 6 \cdot 10^{-7}$ m) is whited-out for visual clarity. The distributions are shown right before the switch to a positive potential, $\tilde{t} = 0^-$, and for different time-periods along the process. System parameters: $L = 100$ nm, $x_c = 0.7$, $W = 5L$, $C_0 = 1$ mM, $V_{Max} = 0.5$ V, $\alpha = -1$, $\delta = 0.5$, and $f = 1$ MHz.

## B. RBIP transport mechanism

To study the transport mechanism of the RBIP we analyze the temporal changes in electric potential, ion concentration, and ion flux, along a time-period in steady state, as shown in FIG. 3**Error! Reference source not found.**. The term steady state here refers to a state in which the initial condition of the device no longer affects its output. The device geometry, and all system parameters are as in FIG. 2(a), except the frequency, which is $f = 100$ kHz. An open system is used with no concentration gradient between the reservoirs ($C_R = C_L = 1$ mM). FIG. 3(a) shows snapshots of the ratchet region at a few key times along the period. The electric potential is illustrated by the dotted black line (left vertical axis). At $\tilde{\tau} = 0^-$, just before the first potential switch, cations (dot-dash red line, right vertical axis) are accumulated at the potential minimum induced by the middle electrode, and anions (solid blue line, right vertical axis) are accumulated at the left and right electrodes, which are the potential maxima. Focusing first on the anions in the right section of the ratchet and right reservoir. Shortly after the switch ($\tilde{\tau} = 0.0025$) most anions drift towards the middle electrode, however, there is a small subset of anions that remain in the right reservoir. The blue shaded area marks this group, and it consists of all the anions in the right reservoir that are in access to the baseline (steady state) distribution (reached at $\tilde{\tau} = 0.5$). In the time interval $0.03 < \tilde{\tau} < 0.5$, most of those anions diffuse to the right. This is because they face a local potential barrier at the right electrode ($\sim 5 - 5.6\ V_{th}$), while towards the bulk of the right reservoir there is no opposing electric field. In the second half period ($0.5 < \tilde{\tau} < 1$), the anions that accumulated near the middle electrode split into two groups, with the majority drifting back towards the right electrode. It is found that more anions are transported back to the right electrode than were extracted from it in the first half of the period. The difference is exactly the number of ions that were transported to the right reservoir in the first part of the cycle. We note that in the first half period, there is also a small subset of anions that remain in the left reservoir. However, because of the spatial asymmetry, the potential barrier at the left electrode is much smaller than it is in the right electrode ($\sim 2.9 - 3.7\ V_{th}$ between $0.03 < \tilde{\tau} < 0.5$). As a result, diffusion towards the left reservoir is much less pronounced compared to the diffusion process towards the right reservoir discussed above. Hence, by $\tilde{\tau} = 0.5^-$, more anions from the left reservoir make the way to the middle electrode, then will go back to the left from the middle electrode in the second half period (see discussion in the next paragraph). Considering that anions and cations in this case have the same diffusivity and that the input signal is temporally symmetric, the same process described above takes place for cations, with a shift in time of half a period. The result is a net flux of both anions and cations in the positive $x$ direction. This is further exemplified in FIG. 3(c), in which the full length of the system is displayed. Far from the ratchet the ion distribution is linear with a diffusion flux ($\propto \partial C / \partial x$) that is equal at both reservoirs.

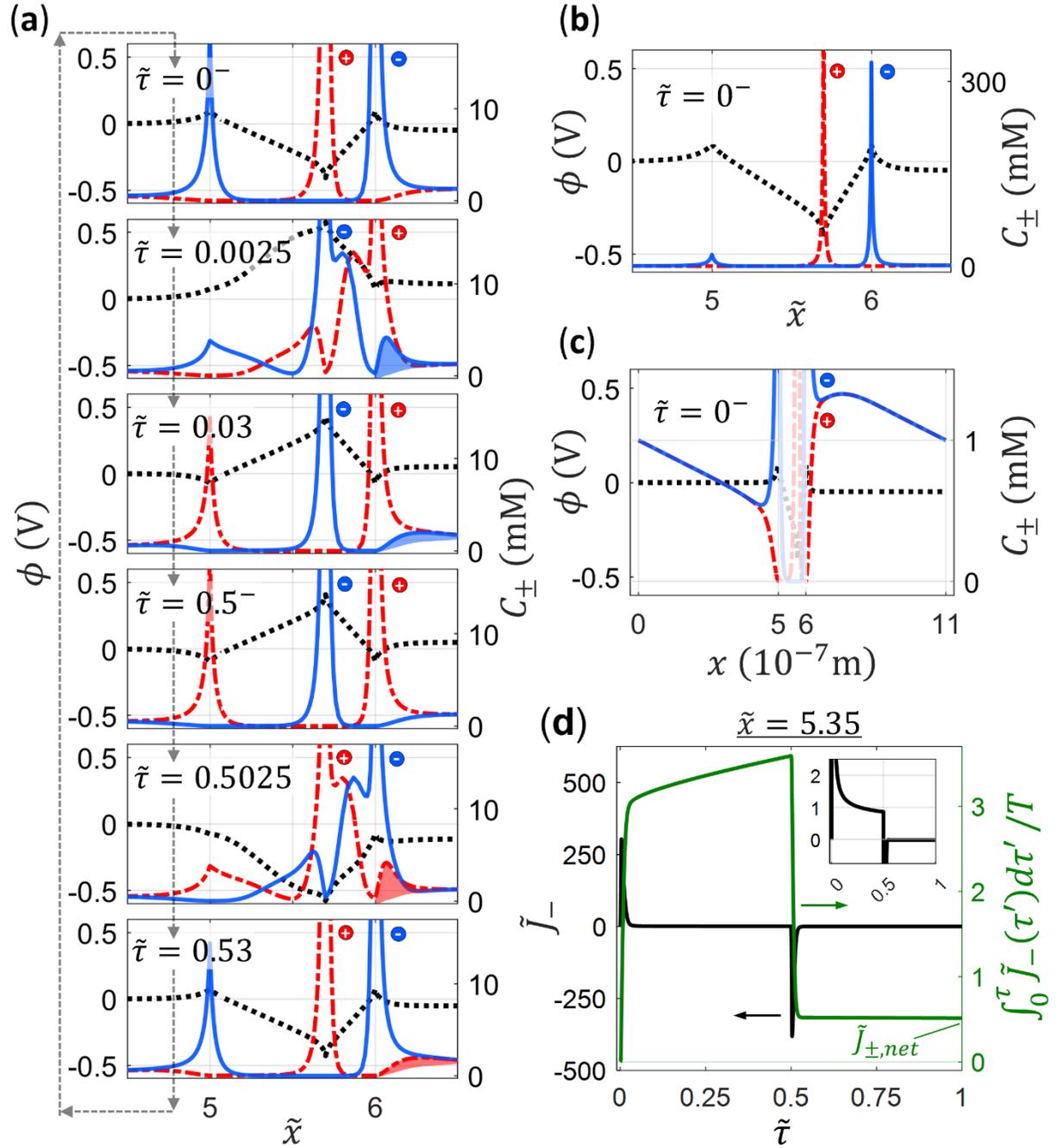

FIG. 3. Analysis of the RBIP transport mechanism. (a) Zoom-in on spatial distribution of cations (dot-dash red line), anions (solid blue line), and electric potential (dot-dot black line), at a few key times along a period. Shaded areas represent ions that are diffusing to the right reservoir. The same results are presented in (b) showing the peak concentration at the electrodes, and in (c) for the full $x$-axis, showing the distribution in the reservoirs. The region inside the membrane ($5 \cdot 10^{-7} < x < 6 \cdot 10^{-7}$ m) is whited-out for visual clarity. (d) Normalized anion flux, $\tilde{J}_-$, inside the membrane at $\tilde{x} = 5.35$ (left $y$-axis), and the integral of $\tilde{J}_-$ up to point $\tau$ along a time-period (right $y$-axis). The inset shows the same data focusing on the slower current components. Results are for an *open system* in steady state. System parameters: $L = 100$ nm, $x_C = 0.7$, $W = 5L$, $D_\pm = 2 \cdot 10^{-9}$ m$^2$/s, $C_0 = 1$mM, $C_R/C_L = 1$, $V_{Max} = 0.5$ V, $\alpha = -1$, $\delta = 0.5$, and $f = 100$ kHz.

FIG. 3(d) shows the normalized anion flux, $\tilde{J}_-$ at the center of the left ratchet section ($\tilde{x} = 5.35$), over the course of a time-period at steady-state (left *y*-axis). It also shows the cumulative flux along a time-period, $\int_0^\tau \tilde{J}_-(\tau')d\tau'/T$ (right *y*-axis). This value is the total contribution to the flux up to a point $\tau$ in the period. It is normalized by $T$, so the result at $\tau = T$ is exactly the net flux, $\tilde{J}_{\pm,net}$, which in steady state is equal everywhere in the system. Only the anion flux is shown, however, the net flux of anions and cations is equal. The flux peak between $\tilde{\tau} = 0$ and $\tilde{\tau} \sim 0.03$, corresponds to anions discharging from the left electrode, drifting along the left ratchet section, and charging the middle electrode. Between $\tilde{\tau} = 0.5$ and $\tilde{\tau} \sim 0.53$, there is an opposite negative flux peak, due to anion motion back toward the left electrode. These fast back-and-forth transport events almost entirely cancel out in terms of their contribution to the net flux. The inset shows the normalized temporal flux focusing on the slower flux components. Although there is an initial sharp decline in its magnitude, the flux is not diminished by $\tilde{\tau} = 0.5$. This non-zero flux is comprised of anions from the left reservoir, which diffuse over the small potential barrier at the left electrode and then drift towards the middle electrode. This is also the reason for the moderate but overall significant increase in the cumulative flux between $0.03 < \tilde{\tau} < 0.5$. There is no opposite transfer of anions to the left between $0.53 < \tilde{\tau} < 1$, due to the large potential barrier at the middle electrode. Thus, the net flux through the left section of the ratchet is mostly a result of a relatively slow transport of ions from the left reservoir between $0.03 < \tilde{\tau} < 0.5$.

The last two paragraphs have described separately the process of ion transport from the left reservoir up to the middle electrode, and from the middle electrode to the right reservoir. The system reaches a steady state when the ion transport rate of both processes is equal. The underlying reason that allows both processes is ultimately derived from the spatial asymmetry of the device, that results in a higher ion concentration and larger potential barrier at the right electrode compared to lower ion concentration and smaller potential barrier at the left electrode, as shown for example at $\tilde{\tau} = 0^-$ in FIG. 3(b). Further discussion on this effect can be found in the Supplementary Material [27].

The potential distribution in the ratchet ($5 < \tilde{x} < 6$ in FIG. 3(a)) is close to a linear saw-tooth at any point during the cycle since there is very little potential screening at the electrodes. This is partly explained by the potential switching times, that are too short for the system to reach the final ion distribution. However, this is not the main reason, since the signal frequency is not very large compared to the fast transport processes (after each potential switch). For example, potential screening does not become significant even for a signal frequency as low as 5000 Hz. Instead, it is mainly the result of the low total amount of ions available within the ratchet area, due to the small length scale of the device and the low bulk concentration. On other cases where the bulk concentration is higher or the device length scale is larger, such as presented in Fig. S1 in the Supplemental Material [27], potential screening is much more pronounced even at higher normalized frequencies. Nevertheless, as will be shown in the next section, minimal potential screening is neither a prerequisite nor is it necessarily an optimal condition for ion pumping.

Ambipolar transport was previously identified, but only for non-symmetric electrolytes ($D_+ \neq D_-$) and a particular subset of input signals [22]. With the current model, that accounts for the coulombic interaction between ions and reservoir effects, ambipolar transport is inherent and it is highly robust. It persists for symmetric (FIG. 2(a), FIG. 3) and non-symmetric electrolytes (FIG. 2(b)); as well as for time-symmetric (FIG. 2(a)-(b), FIG. 3) and non-time-symmetric input signals (see Fig. S2 in the Supplemental Material [27]). In *On-Off* ratchets, where $\alpha = 0$, ambipolar pumping is still achieved. However, due to counter ion interactions, the net flux is much lower, and ion pumping is in the opposite direction than predicted by other models [20,24].

The analysis in this section provides a general understanding of the different time scales that are involved in the temporal evolution of ionic flux. These time scales are consistent with the general dynamics of microelectrochemical system [29]. They include fast processes occurring within the EDLs and slower processes within the bulk of the membrane and the reservoirs, which are associated with large applied voltages (relative to $V_{th}$). The overall operation of the device is a result of a complex interaction between these different processes, which may vary for different electrolyte compositions, system geometries, and input signals. The next section will examine the RBIP performance as a function of a few key parameters of the system – the input signal frequency and amplitude, the ratchet geometry (device length and symmetry factor), and the bulk concentration.

### C. RBIP performance analysis for key parameters

A few key parameters of the RBIP system are analyzed here to determine their effect on pumping performance. FIG. 4(a)-(b) show the normalized net salt flux as a function of normalized signal frequency for different ratchet geometries and bulk concentrations when there is no concentration difference between the reservoirs ($C_R/C_L = 1$). The frequency is normalized by a characteristic time scale for each half of the period and is defined as $\tilde{f} = f/(D/(2L^2))$. Each curve has an optimal frequency where the flux is maximal, which is expected in ratchet systems. FIG. 4(c)-(d) show the normalized net salt flux as a function of the concentration ratio between the reservoirs $C_R/C_L$ at specific signal frequencies. The $x$-axis intersection, marked $(C_R/C_L)_{Max}$, is the maximal concentration ratio that the ratchet can overcome (with zero flux), similar to an open circuit voltage of a solar cell, or to a maximal pressure of a pump. The $y$-axis intersection, marked $\tilde{J}_{Max}$, is the maximal normalized net flux that is achieved when there is no opposing concentration difference, similar to a short-circuit current. At the operating points along the curve, the ratchet is performing work, i.e., it drives an ionic current up a concentration gradient. For comparison a 'no ratchet' curve is shown which is calculated as the diffusion flux for a system of the same total length $2W + L$, and no ratchet present.

FIG. 4(a) and FIG. 4(c) explore the pumping characteristics for a spatial period $L = 0.1$ μm and a bulk concentration $C_0 = 1$ mM. For a spatial asymmetry of $x_C = 0.7$, driving the ratchet at the optimal normalized frequency $\tilde{f} = \tilde{f}_{opt} = 1$ (which corresponds to $f = 100$ kHz), results in a higher pumping performance than $\tilde{f} = 10$ ($f = 1$ MHz) for all concentration ratios. A larger asymmetry factor $x_C = 0.9$ results in a higher optimal frequency and significantly improved pumping performance for all concentration ratios. The last curve noted "Thick Electrodes" is calculated assuming electrodes with a finite thickness of 10 nm. In this case, only the insulators' size is used to calculate the ratchet length, $L$, and asymmetry factor $x_C$. The optimal frequency and pumping performance are somewhat reduced, but the overall behavior is similar.

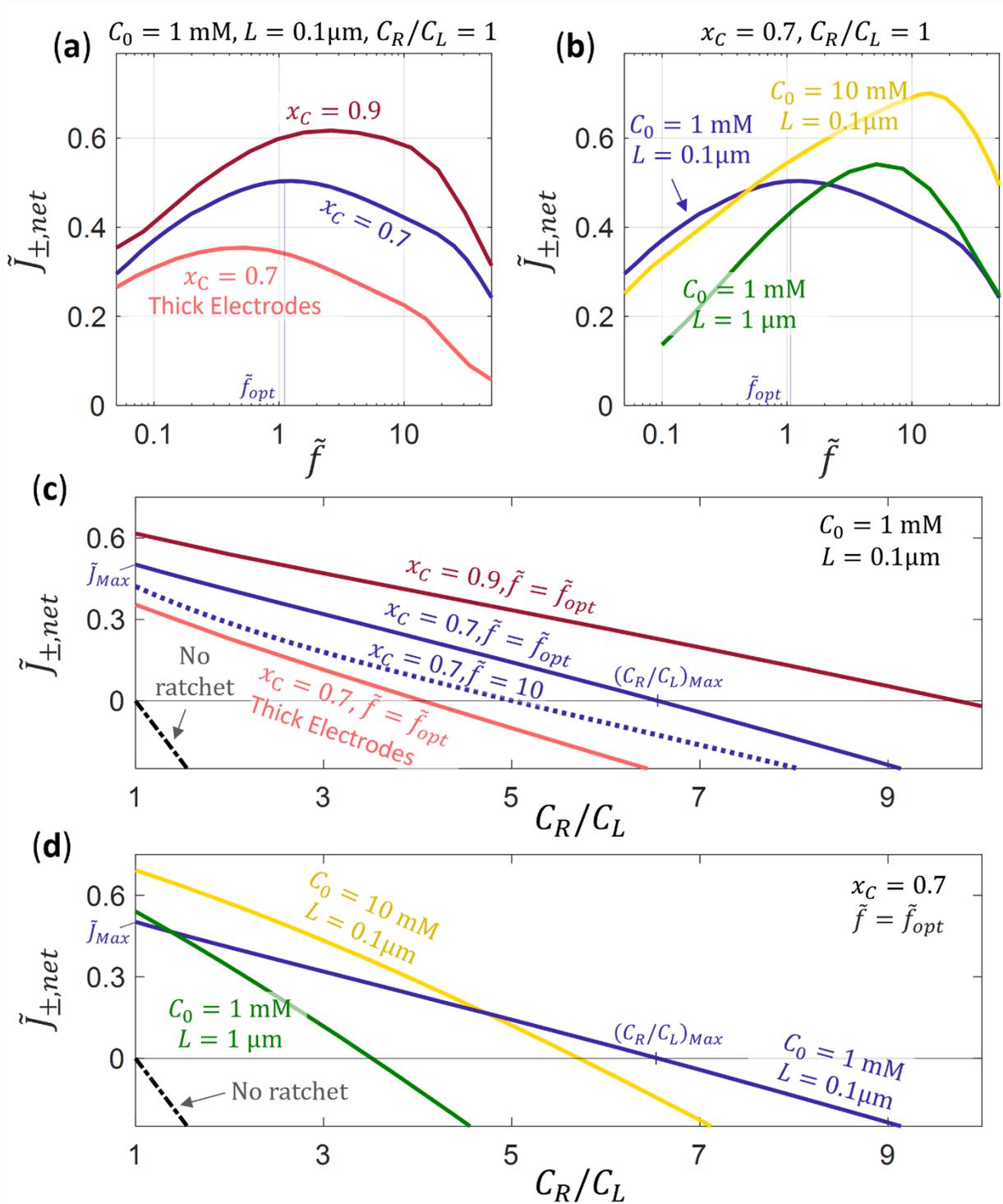

FIG. 4. RBIP performance analysis for different frquencies, ratchet geometries, and bulk concentrations. Normalized net salt flux as a function of (a), (b) normalized signal frequency when $C_R = C_L$, and (c), (d) opposing concentration ratio. General system parameters: $W = 5L$, $D_\pm = 2 \cdot$

$10^{-9}$ m²/s, $C_L = C_0$, $V_{Max} = 0.5$ V, $\alpha = -1$, $\delta = 0.5$. The PNP equations are valid for frequencies near and above the optimal frequency of each presented case.

For a ratchet length $L = 0.1$ μm, when $C_0$ increases from 1 mM to 10 mM the maximal normalized net flux increases by 40% (FIG. 4(b)), which corresponds to an absolute net flux increase by a factor of 14. Part of this increase in absolute flux is due to the 10-fold rise in bulk concentration, but the additional increase in normalized flux for $C_0 = 10$ mM is due to a larger fraction of ions (with respect to the bulk concentration) that are transferred across the membrane on each time-period. This fraction is larger for $C_0 = 10$ mM at any specific frequency above $\tilde{f} = 0.5$, and especially for $\tilde{f} > 10$, as shown in Supplemental Material Fig. S3 [27]. An analogy can be drawn between this and a simple RC circuit, where a higher bulk concentration results in faster transients ($\tau_{RC} = R_b C_D \propto 1/\sqrt{C_0} \propto \lambda_D$, where $R_b$ is the bulk electrolyte resistance, and $C_D$ is the double layer capacitance), which enables operation at higher frequencies. FIG. 4(d) shows that the maximal concentration ratio for $C_0 = 10$ mM is slightly lower than for $C_0 = 1$ mM. However, the maximum normalized output power, which is found by $\left(\tilde{J}_{\pm,net} \ln C_R/C_L\right)_{Max}$, is higher by 35% for $C_0 = 10$ mM. The RBIPs ability to pump both ions against a concentration gradient at moderate salinity levels implies that it can be suitable for brackish water desalination. To achieve the required concentration gradients, a multi-stage process may be needed. A detailed analysis of the pumping efficiency and comparison to other desalination methods is left for future work.

The Poisson-Nernst-Planck (PNP) equations do not consider steric effects [30]. Thus, for larger bulk concentrations, and relatively low signal frequencies, the peak concentration at the electrodes might exceed realistic values. However, it was verified that the ion concentration near (and above) the optimal frequency do not exceed the steric limit (calculated to be 4.8 M, assuming an effective ion size of 7 Å for both cations and anions [30]). The performance of the device is highly dependent on the charging/discharging properties of the EDLs, which are in turn highly effected by steric effects near the electrodes. It was also shown that under relatively large voltages, the charging/discharging behavior of the EDLs is non monotonic with voltage amplitude [30]. Therefore, predicting the impact of steric effects at higher salinities and/or lower frequencies is not trivial and will require further investigation that is beyond the scope of this work.

FIG. 4(b) and FIG. 4(d) also show curves for a device with a spatial period of $L = 1$ μm, and a bulk concentration $C_0 = 1$ mM. A larger ratchet length requires a longer ion transport time, which naturally results in a lower optimal frequency, $f_{opt} = 5$ kHz ($\tilde{f}_{opt} = 5$). This also leads to a lower (absolute) net salt flux compared to a smaller ratchet length. The potential distribution and ion concentration for such a device, working at the optimal frequency, are shown in Supplemental Material Fig. S1 [27]. Unlike the 100 nm long device, here the total number of ions within the membrane are enough to completely screen the applied potential (except for very short durations shortly after potential switches). Nevertheless, the normalized pumping performance is only somewhat reduced, resulting in a lower maximal concentration ratio, and the overall transport mechanism is similar to that shown in FIG. 3. Supplemental Material Fig. S4 [27] compares the operation of this device ($L = 1$ μm, $C_0 = 1$ mM) to a device with the same ratio of $\lambda_D/L$, i.e., a ratchet length $L = 0.1$ μm, and a bulk concentration $C_0 = 100$ mM. Although the absolute net flux and operational frequency range vary by orders of magnitude between the cases, the normalized curves perfectly coincide. This demonstrates the key importance of this ratio to the device pumping characteristics. For $C_0 = 100$ mM the steric concentration limit is exceeded even near and above the optimal frequency, thus this result only serves to exemplify the effect of the $\lambda_D/L$ ratio. Perhaps contrary to intuition, the best pumping performance is not obtained when the Debye length approaches the length scale of the device, $L$ (i.e., when potential screening by the electrolyte is negligible), but there is an optimal $\lambda_D/L$ value, which can be much smaller than one. For the ratchet parameters in FIG. 4(b) and FIG. 4(d), the optimal ratio is in the range $0.1 < (\lambda_D/L)_{opt} < 0.01$. Thus, when designing a device, it is important to consider that although a smaller device length scale, $L$, increases the net (absolute) salt flux, it also increases $\lambda_D/L$, which might somewhat reduce the pumping efficiency.

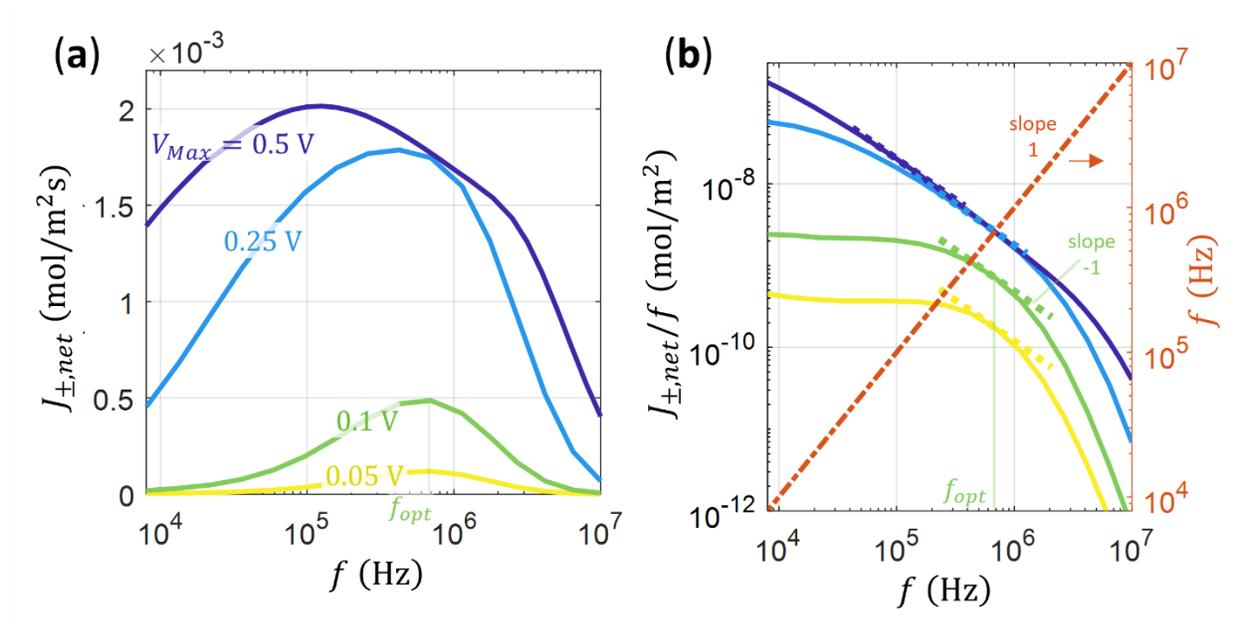

FIG. 5. Net salt flux, $J_{\pm,net}$ (a), and mols transported per time-period, $J_{\pm,net}/f$ (b), as a function of signal frequency for different signal amplitudes. The dotted lines in (b) show the slope (in log-log) of the curves at the optimal frquency, and for reference the curve $f = f$ (right y-axis, dash-dot line). General system parameters: $L = 100\ nm$, $x_C = 0.7$, $W = 5L$, $D_\pm = 2 \cdot 10^{-9}\ m^2/s$, $C_0 = 1mM$, $C_R/C_L = 1$, $\alpha = -1$, $\delta = 0.5$.

Consideration should be given to the optimal frequency (at which the net ion flux is greatest). Predicting the optimal frequency for different ratchet geometries and bulk concentrations is not trivial. There are multiple time scales governing ion transport, and they depend on the length of the different sections in the system, and on the Debye length ($\propto 1/\sqrt{C_0}$) [29]. It also varies as a function of input signal amplitude, $V_{Max}$. FIG. 5(a) shows the calculated net salt flux as a function of the input signal frequency and several input signal amplitudes. FIG. 5(b) shows the number of mols transferred across the membrane per time-period, $J_{\pm,net}/f$, calculated with the same data. For $V_{Max} = 0.05$ V and $V_{Max} = 0.1$ V, at the lower frequencies up to ~100 kHz, $J_{\pm,net}/f$ is almost constant since the system reaches equilibrium between each potential switch. Increasing the ratchet frequency leads to more transport cycles per unit time, thus resulting in a linear increase in net flux. As the frequency increases above ~100 kHz, $J_{\pm,net}/f$ starts to decrease. However, since this decrease is moderate at first, the contribution of the added transport cycles with frequency is more dominant and thus the net flux continues to increase. The optimal frequency is achieved at the point where the decrease in $J_{\pm,net}/f$ starts to outweigh the contribution of the added transport cycles. On the log-log graph in FIG. 5(b) this is equivalent to a slope that is more negative than -1. As $V_{Max}$ increases, $J_{\pm,net}/f$ also increases, and as a result so does the net flux. Interestingly, the optimal frequency decreases as a function of voltage amplitude. This may seem surprising at first since larger electric fields lead to faster ion transport, and indeed this is the case for the initial response of the system to each potential switch. However, when high voltages are applied (relative to $V_{th}$), the slower transport processes become more dominant (as discussed in section III.B). This results in longer times to reach equilibrium, and a significant increase in $J_{\pm,net}/f$ at the lower frequencies, thereby reducing the optimal frequency. Fig. S5 in the Supplemental Material [27] shows the maximal net flux (for $C_R/C_L = 1$) and the maximal concentration ratio, $(C_R/C_L)_{max}$, as a function of $V_{Max}$ for $f = 100$ kHz. The net flux increases as the effect of long time-scale transport processes becomes more and more significant, until reaching saturation above $V_{Max} = 0.4$ V. The maximal concentration ratio follows a similar trend, except the increase with amplitude only starts to become more moderate at $V_{Max} = 0.4$ V. Due to steric concentration limits, a study of the effects of higher signal amplitudes is left for future work. The increase of the maximum flux and $(C_R/C_L)_{max}$ with $V_{Max}$ and their saturation at a higher voltage implies that there is an amplitude for which the pumping efficiency is optimal. Since the thermodynamics of this process are out of the scope of this work, a detailed examination of this effect is left for future study.

An additional local frequency optimum may exist for non-symmetric electrolytes ($D_+ \neq D_-$), for input signal parameters that produce ambipolar transport in a ratchet model that does not account for columbic interactions [22]. For example, Fig. S6 in the Supplemental Material [27] shows the case for NaCl. It can be seen that [22], there is a local increase in net flux compared to a symmetric electrolyte ($D_\pm = 2 \cdot 10^{-9}$ m²/s) under the same conditions. As predicted by the noninteracting model, at the local frequency optimum cations and anions are simultaneously driven in the same direction and with the same velocity, therefore reducing the impeding effect of columbic interactions within the membrane.

## IV.  CONSLUSIONS

We have simulated the performance of an electrically activated ratchet-based ion pump (RBIP) membrane in an electrolytic environment accounting for coulombic interaction between ions. The study has shown that this device drives both cations and anions in the same direction (ambipolar pumping or salt pumping) and up a concentration gradient. This finding was highly robust for many electrolyte compositions and input signals. The results show that with an input signal amplitude of 0.5 V, the RBIP can pump ions against opposing concentration ratios as high as 10:1, without the need for electrochemical reactions at the electrodes. This work is a first step in the study of multi-layered RBIPs, and it provides the basis for understating of the ambipolar pumping mechanism, and its possible applications. The multi-layered architecture holds great promise for fabricating membranes that can realize different electric potential landscapes within a solution, potentially enabling high efficiency water desalination.

## ACKNOWLEDGMENTS

GS thanks the Azrieli Foundation for financial support within the Azrieli Fellows program. This work is partially funded by the European Union (ERC, ESIP-RM, 101039804).

The authors declare no competing financial interest.

*Large Applied Voltages. I. Double-Layer Charging*, Phys. Rev. E **75**, 021502 (2007).

**Supplemental Material for:**

# Ambipolar ion pumping with ratchet-driven active membranes

Alon Herman[1] and Gideon Segev[1*]

[1]School of Electrical Engineering, Tel Aviv University, Tel Aviv 6997801, Israel

[*]email: gideons1@tauex.tau.ac.il

**Electric Potential Boundary Conditions**

As discussed in the main text, a reference point for the electric potential was defined at the left edge of the system (Eq. (3c)), while the right edge was defined as electrically insulating (Eq. (3d)). For complete symmetry between the two reservoirs, the left edge should also be defined as electrically insulating in addition to the fixed potential, however, this is not allowed by the COMSOL user interface. Nevertheless, as was verified by examining the data, this has no practical importance for almost all the simulations in this work, where a binary and symmetric electrolyte ($D_+ = D_-$) was used. Since the outer edges are electroneutral, when the diffusion coefficient of Anions and Cations is equal, there is in any case no reason for an electric field to develop at both edges. For a *closed system*, since there is no ionic flux at the outer edges, the electric field is also zero even if $D_+ \neq D_-$, and this was also verified by examining the data of the NaCl simulation shown in Fig 2(b). For NaCl in an *open system* (Figure S5) a small electric field may develop at $x = 0$. Thus, the reference point for the electric potential was moved to the outer electrodes and an electrical insulation condition was applied at the left outer edge. The modified potential boundary conditions for NaCl in Figure S5 are thus:

$$\phi(x = W, t) = \phi(x = W + L, t) = 0, \tag{S1a}$$

$$\phi(x = W + x_c L, t) = V_{in}(t), \quad \text{(S1b)}$$

$$\partial\phi/\partial x\,(x = 0, t) = \partial\phi/\partial x\,(x = 2W + L, t) = 0. \quad \text{(S1c)}$$

A simulation with these boundary conditions was conducted for the symmetric electrolyte ($D_+ = D_-$) and was shown to produce the same results as with the potential boundary conditions that were described in the main text (eq. 3a-d).

**Net Flux Calculation for a Closed System**

In a closed system the ions' flux is set to zero at the outer edges of the left and right reservoirs, and as a result, the flux is not spatially uniform in the reservoirs. The net flux (per period) could be found by averaging over the entire $x$ axis domain. However, since there are very fast changes in ion concentration and electric potential at the ratchet area, to get an accurate result this requires an extremely high time resolution, which is computationally demanding. Instead, an approximate calculation was used which is based on the time derivatives of ions' concentration at the reservoir's outer edges, $C_R$ and $C_L$. Since these values have a much slower response to the ratchet operation, a significantly lower time resolution could be used. It is assumed that the concentration is uniform in the reservoirs, and thus the average flux in each reservoir is calculated as $J_{\pm,R} = J_R \approx W(\partial C_R/\partial t)$ and $J_{\pm,L} = J_L \approx -W(\partial C_L/\partial t)$. The net flux of the system is then taken as an average of the net flux in both reservoirs, according to $J_{\pm,net} = (J_{R,net} + J_{L,net})/2$.

**Initial Conditions for an Open System**

In an open system, the initial conditions for the electric potential and ions concentration are calculated by solving a preliminary stationary problem ($\partial C_\pm/\partial t = 0$). It uses the same boundary conditions as described in the main text, except the voltage input is constant in time, and equal to the time average of the signal $V_{in-const} = \delta V_{Max} + (1-\delta)\alpha V_{Max}$.

**Additional Simulation Definitions**

The coupled Nernst – Planck – Poisson equations were solved using COMSOL Multiphysics® v6.1. The spatial discretization used quadratic Lagrange elements. The one-dimensional mesh was defined by an exponential growth in element size away from each of the three electrodes, with at least 10 elements within 1nm from each side of the electrodes. The time-steps were automatically generated by the solver, using COMSOL's *events* interface to define the abrupt and periodic changes in electric potential input (at the electrodes). The convergence criterion for the estimated error in each time step used a relative tolerance of 0.001 and an absolute tolerance of 0.0001.

**Higher Ion Concentration at the Right Electrode**

Spatial asymmetry is essential for ratchet driven transport. This asymmetry results in: (i) a higher ion concentration and larger potential barrier at the right electrode that reduces the diffusion of ions from the right reservoir back to the middle electrode, and (ii) a lower ion concentration and smaller potential barrier at the left electrode that allows for the injection of ions from the left reservoir to the middle electrode (FIG. 3(b)). This difference between ion concentration at the left and right electrodes is a result of a fast transport event that takes place near the middle electrode shortly after each potential switch. As shown in FIG. 3(a), shortly after the second switch ($\tilde{\tau} = 0.5025$) a significantly larger portion of anions are transported to the right electrode, than the anions that are driven to the left electrode. The main reason is the higher electric field in the right section (due to the shorter section length). During this short period ($0.5 < \tilde{\tau} < 0.5025$), some of the anions that were to the left of the middle electrode at $\tilde{\tau} = 0.5$ diffuse over the potential barrier and then drift to the right. This transport mechanism is similar to the phenomenon that was termed *injection* in a previous work [1].

**Supplemental Material Figures**

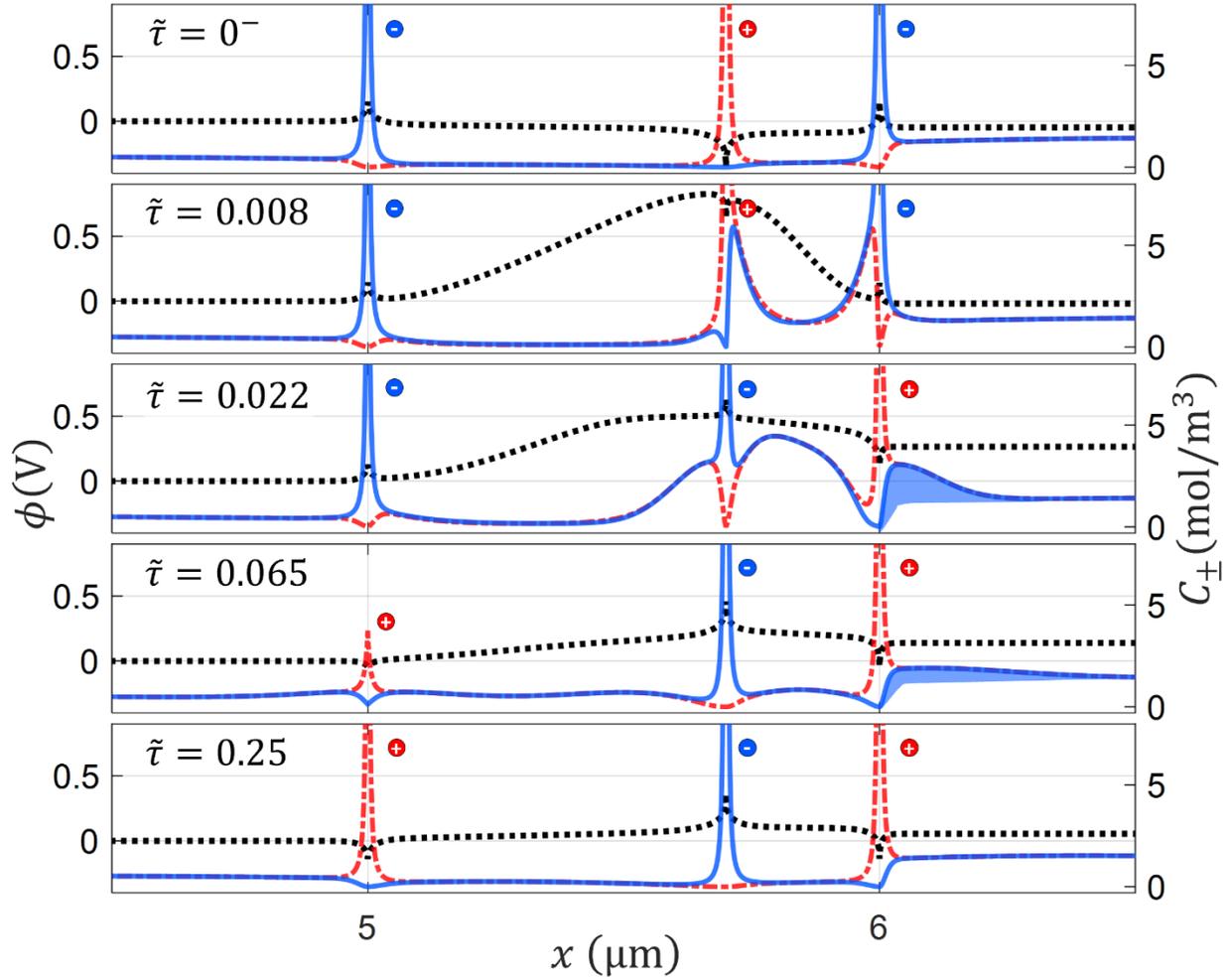

FIG. S1. Spatial distribution of cations (dot-dash red line), anions (solid blue line), and electric potential (dot-dot black line), at a few key times along a period. Partial $x$-axis is presented around the ratchet area. The net ion flux is $2.2 \cdot 10^{-4}$ mol/m²s. Shaded areas represent ions that are diffusing to the right reservoir. Results are for an open system in steady state. System parameters: $L = 1$ μm, $x_C = 0.7$, $W = 5L$, $D_\pm = 2 \cdot 10^{-9}$ m²/s, $C_0 = 1$mM, $C_R/C_L = 1$, $V_{Max} = 0.5$ V, $\alpha = -1$, $\delta = 0.5$, and $f = f_{opt} = 5$ KHz.

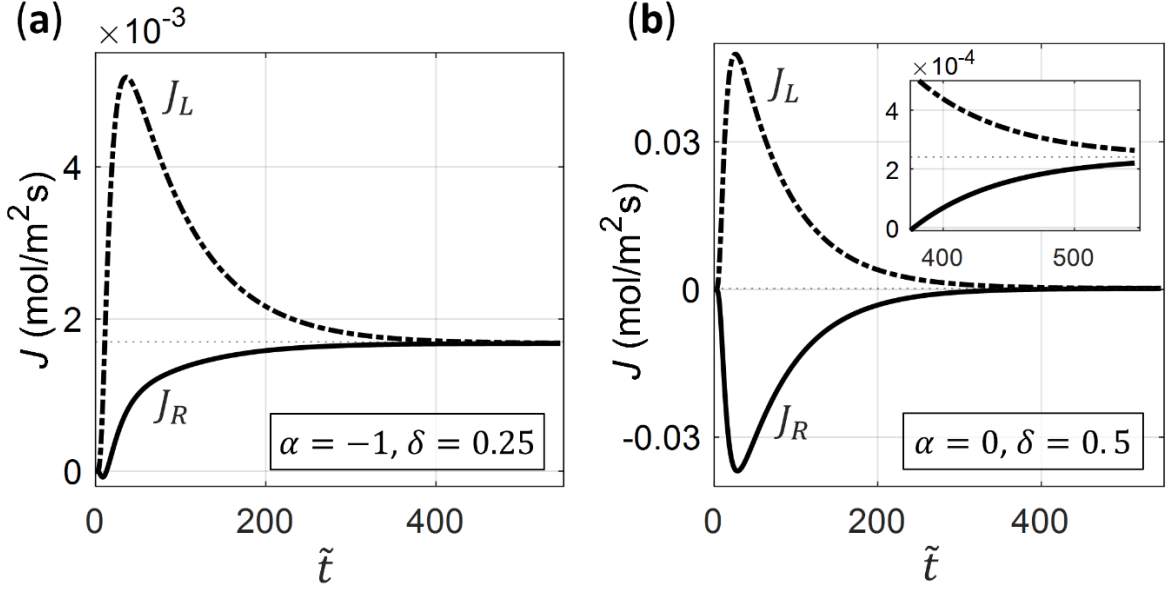

FIG. S2. RBIP operation for non-time-symmetric input signals. Net salt flux over time in an open system, at the left reservoir edge ($J_L$, dot-dash line) and the right reservoir edge ($J_R$, filled line), for (a) $V_{Max} = 0.5$ V, $\alpha = -1$, $\delta = 0.25$, and (b) $V_{Max} = 1$ V, $\alpha = 0$, $\delta = 0.5$. The inset in (b) zooms-in on the final periods. System parameters: $L = 100$ nm, $x_C = 0.7$, $W = 5L$, $C_L = C_R = C_0 = 1$mM, $D_\pm = 2 \cdot 10^{-9}$ m$^2$/s, and $f = 1$ MHz.

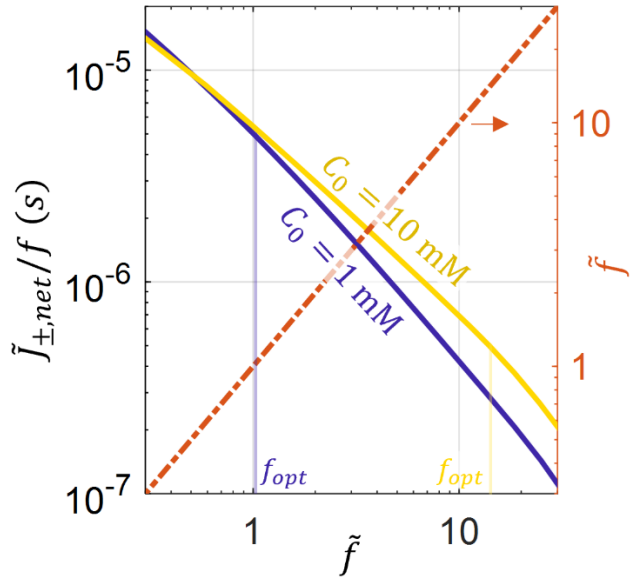

FIG. S3. Normalized number of mols transported across the membrane per time-period, $\tilde{J}_{\pm,net}/f$, as a function of normalized signal frequency for $C_0 = 1$ mM and $C_0 = 10$ mM. The curve $\tilde{f} = \tilde{f}$ is shown for reference on the right y-axis (dash-dot line). General system parameters: $L = 100$ nm, $x_C = 0.7$, $W = 5L$, $D_\pm = 2 \cdot 10^{-9}$ m$^2$/s, $C_R/C_L = 1$, $\alpha = -1$, $\delta = 0.5$.

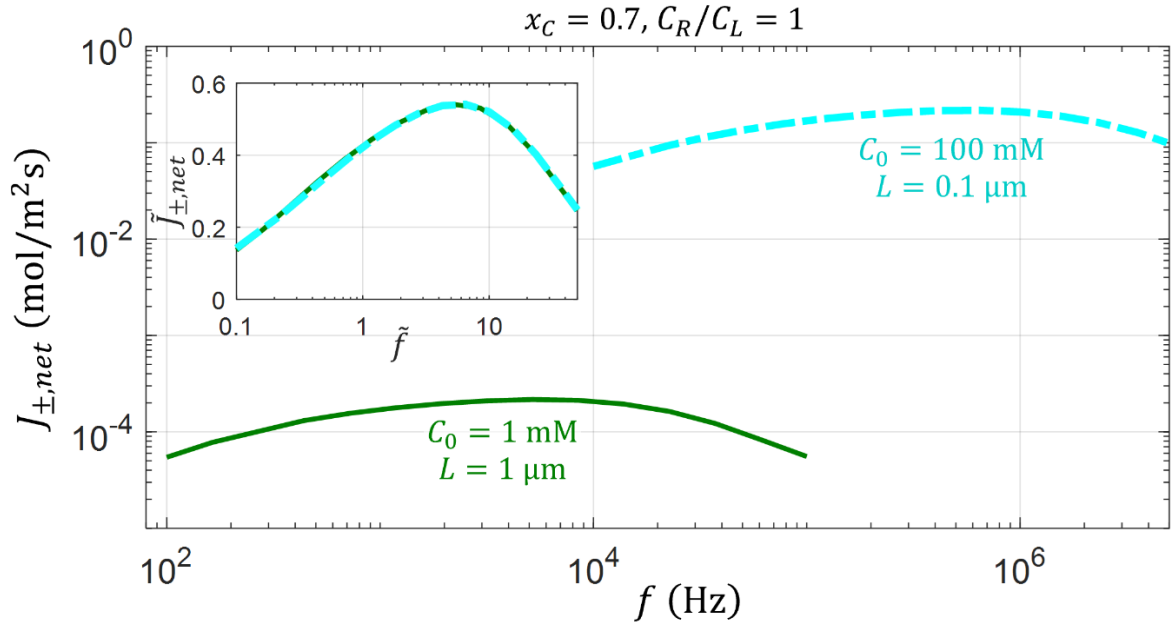

FIG. S4. Net salt flux as a function of signal frequency for two devices with the same ratio of Debye length to device length, $\lambda_D/L$. The inset shows the same data, only for normalized net flux and normalized frequency. Although the absolute net flux and frequency range vary by orders of magnitude between the cases, the normalized curves perfectly coincide. The curve for $L = 0.1$ μm, $C_0 = 100$ mM is given here only as reference, since the steric concentration limit is exceeded for the entire presented frequency range. System parameters: $x_C = 0.7$, $W = 5L$, $D_\pm = 2 \cdot 10^{-9}$ m²/s, $C_R/C_L = 1$, $V_{Max} = 0.5$ V, $\alpha = -1$, $\delta = 0.5$.

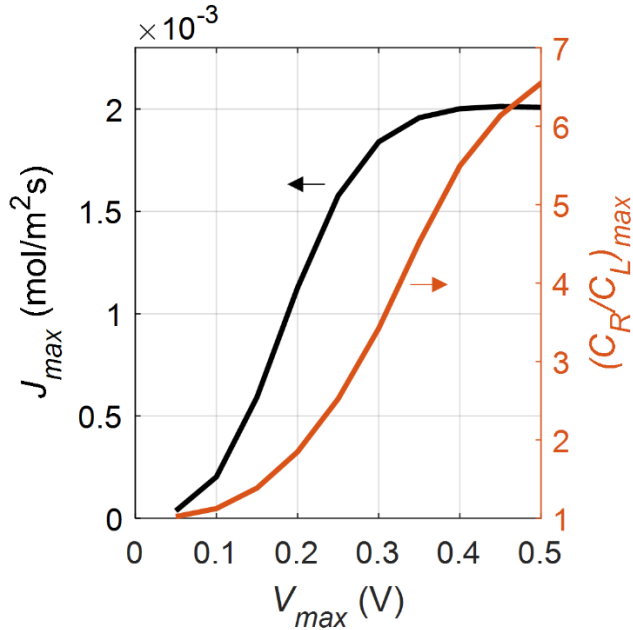

FIG. S5. Maximal net salt flux at $C_R/C_L = 1$ (left y-axis) and maximal concentration ratio (right y-axis) as a function of voltage amplitude. System parameters: $x_C = 0.7$, $W = 5L$, $D_\pm = 2 \cdot 10^{-9}$ m$^2$/s, $C_0 = 1$ mM, $C_R/C_L = 1$, $\alpha = -1$, $\delta = 0.5$, $f = 100$ kHz.

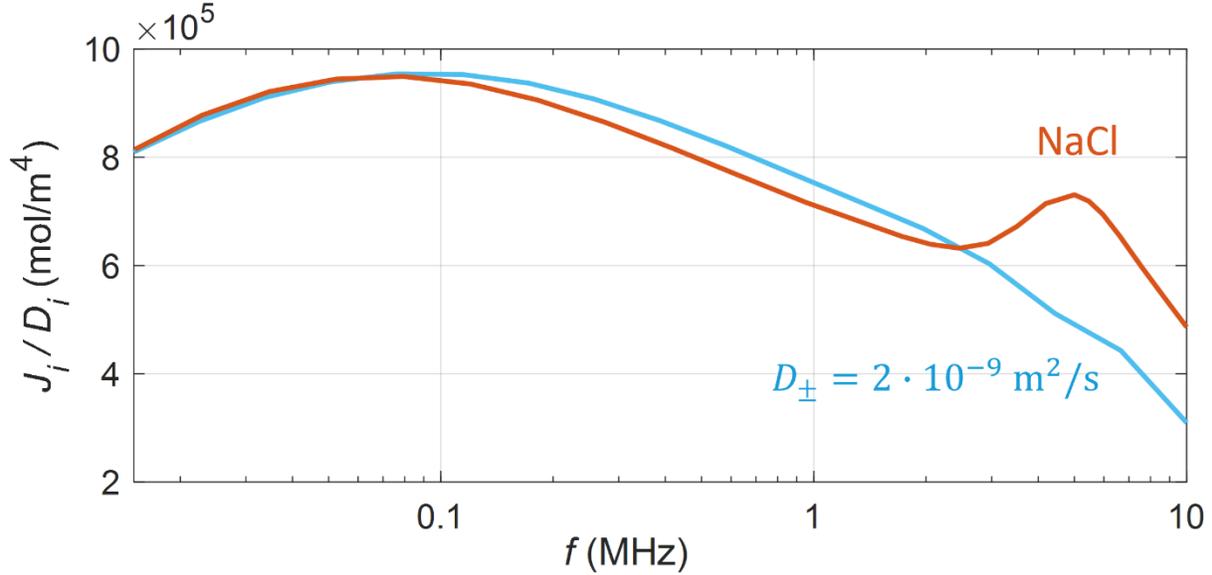

FIG. S6. Net ion flux normalized by diffusion coefficient as a function of signal frequency, for a symmetric electrolyte ($D_\pm = 2 \cdot 10^{-9}$ m$^2$/s) and NaCl ($D_{\text{Na}^+} = 1.33 \cdot 10^{-9}$ m$^2$/s, $D_{\text{Cl}^-} = 2.03 \cdot 10^{-9}$ m$^2$/s). Under these input signal conditions ($\alpha = -0.5$, $\delta = 0.25$), a local frequency optimum exists for NaCl at $f = 5$ MHz. System parameters: $L = 100$ nm, $x_C = 0.7$, $W = 5L$, $C_R = C_L = C_0 = 1$mM, and $V_{Max} = 1$ V. The potential boundary conditions are modified for the NaCl curve as described in the section "Electric Potential Boundary Conditions" of the Supplemental Material.